\documentclass[twocolumn,prl,aps,epsf,showpacs,floatfix]{revtex4}

\usepackage{graphicx}
\usepackage{amssymb}

\begin{document}

\title{Search for Massive Neutrinos in the Decay
 $\pi \rightarrow \mbox{e} \nu$}

\author{M. Aoki$^1$, M. Blecher$^2$, D.A. Bryman$^3$, S. Chen$^4$,
M. Ding$^4$, L. Doria$^5$, P. Gumplinger$^5$, C. Hurst$^3$, A. Hussein$^6$,
Y. Igarashi$^7$, N. Ito$^1$, S.H. Kettell$^8$, L. Kurchaninov$^5$,
L. Littenberg$^8$, C. Malbrunot$^3$, T. Numao$^5$, R. Poutissou$^5$,
A. Sher$^5$, T. Sullivan$^3$, K. Yamada$^1$, M. Yoshida$^1$, D. Vavilov$^5$\\
(PIENU Collaboration)}

\affiliation{
$^1$Physics Department,
Osaka University, Toyonaka, Osaka, 560-0043, Japan\\
$^2$Virginia Tech., Blacksburg, VA, 24061, USA\\
$^3$Department of Physics and Astronomy,
University of British Columbia, Vancouver, B.C., V6T 1Z1, Canada\\
$^4$Department of Engineering Physics,
Tsinghua University, Beijing, 100084, China\\
$^5$TRIUMF, 4004 Wesbrook Mall, Vancouver, B.C., V6T 2A3, Canada\\
$^6$University of Northern British Columbia,
Prince George, B.C., V2N 4Z9, Canada\\
$^7$KEK, 1-1 Oho, Tsukuba-shi, Ibaragi, Japan\\
$^8$Brookhaven National Laboratory, Upton, NY, 11973-5000, USA
}

%\collaboration{PIENU Collaboration}

\date{\today}

%\end{frontmatter}

\begin{abstract}
Evidence of massive neutrinos in the
$\pi^+ \rightarrow \mbox{e}^+ \nu$ decay spectrum was sought with the 
background $\pi^+ \rightarrow \mu^+ \rightarrow \mbox{e}^+$ decay chain highly suppressed.
Upper limits (90 \% C.L.) on the neutrino mixing matrix element $|U_{ei}|^2$
in the neutrino mass region 60--129 MeV/c$^2$ were set 
at the level of $10^{-8}$.
\end{abstract}


\maketitle

\section{Introduction}

A natural extension of the Standard Model (SM)
incorporating neutrino mass and possibly
explaining the origin of dark matter involves the inclusion 
of sterile neutrinos
mixing with the ordinary neutrinos \cite{sterile}.
The weak eigenstates $\nu_{\chi_k}$ of such neutrinos are related to the
mass eigenstates $\nu_i$ by a unitary matrix,
$\nu_{\ell} =\Sigma_{i=1}^{3+k} U_{\ell i} \nu_i$,
where $\ell = e, \mu, \tau, \chi_1, \chi_2...\chi_k$.
An example of a sterile neutrino
model is the Neutrino Minimal Standard Model
that adds to the SM three massive gauge-singlet fermions
(sterile neutrinos) \cite{numsm}.
In the context of this model, 
a search for extra peaks in the $\pi^+ \rightarrow \mbox{e}^+ \nu$ decay spectrum is sensitive
to sterile neutrinos
depending on the mass hierarchy structure and 
choice of parameters
\cite{niigata}.

The decay $\pi^+ \rightarrow \mbox{e}^+ \nu$ ($E_{e^+} = 69.8$ MeV)
with a branching ratio of $R = (1.230 \pm 0.004) \times 10^{-4}$
\cite{oldtriumf,oldpsi,pdg2010}
is helicity-suppressed by $(m_e /m_{\mu})^2$ in the SM.
The relaxation of this condition for massive neutrinos facilitates
the search for extra peaks in the lower positron energy region.
Previous results \cite{oldneutrino}
at the level of $|U_{ei}|^2 < 10^{-7}$
 in the neutrino mass region of
70--130 MeV/c$^2$ were limited by the
presence of unsuppressed $\mu^+ \rightarrow \mbox{e}^+ \nu \overline{\nu}$ decay background 
($E_{e^+} = 0.5 - 52.8$ MeV)
originating
from decay-in-flight of pions \cite{oldtriumf}.
The TRIUMF PIENU experiment \cite{pienu}
aiming at a more precise measurement of
the branching ratio $R$
was designed to further reduce this
background.

In this paper, we present results of the search for low-energy
peaks in the background-suppressed spectrum of $\pi^+ \rightarrow \mbox{e}^+ \nu$
decays at rest.
\\

\section{Experiment}

\begin{figure}[htb]
\centering
\includegraphics*[width=7.5cm]{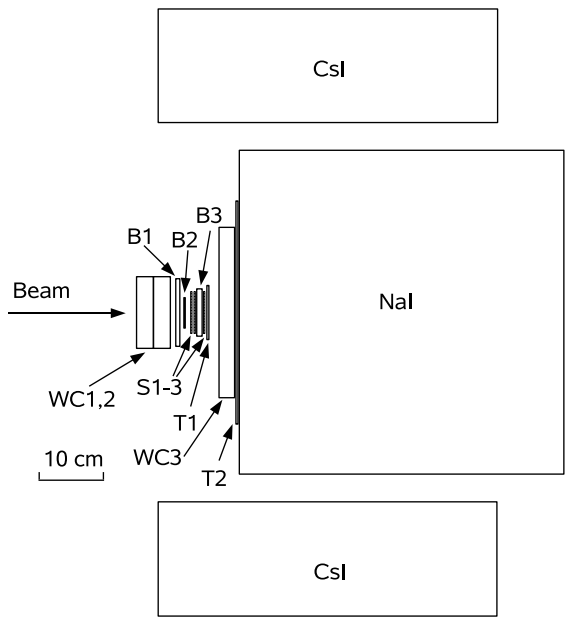}
\caption{PIENU Detector (see text).}
\label{detector}
\end{figure}

The extension of the TRIUMF M13 beam line \cite{m13},
which suppressed positrons in the beam to
$< 2$ \% of pions, delivered a 75$\pm$1 MeV/c pion beam to the experiment.
Figure \ref{detector} shows a schematic view of the detector.
The $\pi^+$ beam was degraded by two thin plastic scintillators
B1 and B2 (6 mm and 3 mm thick, respectively) 
and stopped in an 8-mm thick active target
(B3) at a rate of $5 \times 10^4$ $\pi^+$/s.

In order to obtain a large solid angle for $e^+$ detection from the
decay $\pi^+ \rightarrow \mbox{e}^+ \nu$ while reducing variations
in material along the positron path, the primary calorimeter
was placed on the beam axis.
It consisted of two plastic scintillators
 (3-mm thick T1 and 6-mm thick T2) 
for positron identification and
a 48-cm (dia.) $\times$ 48-cm (length) single-crystal
NaI(T$\ell$) detector \cite{bnl} viewed by 19 photo-tubes.
Two rings of 97 pure CsI crystals \cite{e949}
(9 radiation length radially) surrounded the NaI(T$\ell$) 
crystal to reduce leakage of the electromagnetic shower
produced by a positron.
The observed energy response of the NaI(T$\ell$) detector to
positrons from $\pi^+ \rightarrow \mbox{e}^+ \nu (\gamma)$ decays included a peak at
70 MeV and a low-energy tail distribution.

Pion tracking  was provided by wire chambers (WC1 and WC2) 
each with three-planes
(oriented at 0$^{\circ}$, $\pm 60^{\circ}$) 
 at the exit of the beam line, 
and four planes of 0.3-mm thick X and Y Si-strip counters 
S1X/Y and S2X/Y located immediately
upstream of the B3 counter.
 Positron tracking came from
 one set of X and Y Si-strip counters (S3X and S3Y) immediately
downstream of the B3 counter, and three layers of
wire chambers (WC3) in front of the NaI(T$\ell$) crystal.

A coincidence of beam counters,
B1, B2 and B3, with a high B1 threshold defined
the pion signal, and a coincidence of positron counters, 
T1 and T2, defined the decay-positron
signal. A coincidence of pion and positron signals with a time window
of --300 ns to 500 ns was the basis of the main trigger logic.
This was prescaled by a factor of 16 to form an unbiased trigger 
(Prescaled trigger),
and the early time window (2--40 ns) provided another trigger  
(Early trigger) that
contained most of $\pi^+ \rightarrow \mbox{e}^+ \nu$ decays. Beam positrons for detector calibration
were accumulated by
a separate trigger that required a low pulse height in the B1 counter.
A typical trigger rate was 600 Hz, of which
 240 Hz was for Prescaled, 160 Hz for Early, and
5 Hz for Beam positron.

The pulse shapes from the plastic counters,
the CsI crystals and Si-strip detectors, 
and the NaI(T$\ell$) crystal 
were digitized at 500 MHz, 60 MHz, and 30 MHz, respectively.
\\

\section{Analysis}

\subsection{Event selection}

The charge and time of each pulse were first extracted from the waveform.
Energy calibration of each detector was based on the energy loss of
minimum ionizing particles (75 MeV/c positrons in the beam)
 for plastic and Si-strip counters, and
70 MeV positrons from $\pi^+ \rightarrow \mbox{e}^+ \nu$ decays for the NaI(T$\ell$) signals.
Gain instability was corrected on a run-by-run basis (typically, every
ten minutes).

Events originating from stopped pions were selected 
based on their energy losses
in B1, B2, S1X/Y and S2X/Y, and the time-of-flight
with respect to the primary-proton beam burst (or the phase of the 
cyclotron Radio Frequency).
Events caused by beam positrons and muons were negligible after the cuts.
Any events with extra activity in the beam counters including WC1 and WC2
were rejected. About 40 \% of events survived the cuts.

An energy-loss cut
based on the minimum energy-loss in T1, T2, S3X and S3Y
was used to select decay-positrons. The cut reduced
beam muons directly hitting those counters and protons
coming from pion-nucleus reactions to a negligible level.
 A fiducial cut,
requiring the hit position at WC3 to be within 8 cm from the beam axis,
was imposed to reduce the low energy tail of the $\pi^+ \rightarrow \mbox{e}^+ \nu$ peak
due to shower leakage from the NaI(T$\ell$) crystal.
The low energy tail was further reduced by rejecting events
with energy above 6 MeV in the CsI crystals.

\begin{figure}[htb]
\centering
\includegraphics*[width=8cm]{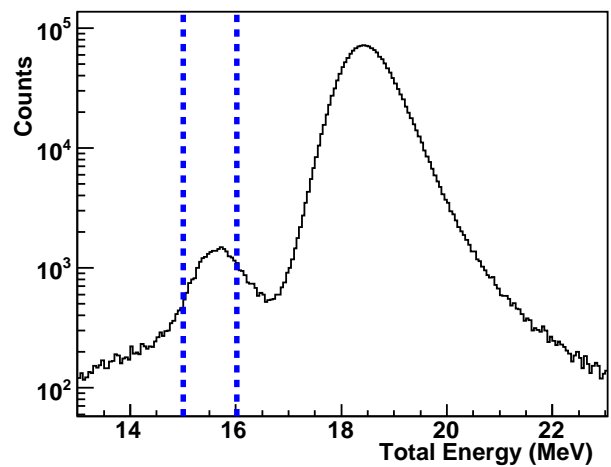}
\caption{Total energy in the beam counters. The peaks at 15.6 and 18.5 MeV
are from $\pi^+ \rightarrow \mbox{e}^+ \nu$ and $\pi^+ \rightarrow \mu^+ \rightarrow \mbox{e}^+$ decays, respectively. The selected region is
between the vertical dashed lines.}
\label{total}
\end{figure}

In order to enhance the small $\pi^+ \rightarrow \mbox{e}^+ \nu$ decay signal,
the $\pi^+ \rightarrow \mu^+ \rightarrow \mbox{e}^+$ ($\pi^+ \rightarrow \mu^+ \nu$ decay followed by $\mu^+ \rightarrow \mbox{e}^+ \nu \overline{\nu}$ decay)
background was suppressed 
using timing cuts to take advantage of
the lifetime difference between pions and muons
($\tau_{\pi} = 26 \mbox{ns}$
and $\tau_{\mu} = 2197 \mbox{ns}$).
The time window selected was 2--33 ns following the
pion stop time.
Since the decay $\pi^+ \rightarrow \mbox{e}^+ \nu$ involves only two charged particles 
 while the decay $\pi^+ \rightarrow \mu^+ \rightarrow \mbox{e}^+$ has
three charged particles with an extra kinetic energy of 
4.1 MeV deposit from the $\pi^+ \rightarrow \mu^+ \nu$ decay in B3,
pulse shape discrimination based on the likelihood for
two and three pulses, and the total energy in the beam
counters were also very effective in $\pi^+ \rightarrow \mu^+ \rightarrow \mbox{e}^+$ background suppression.
 In Figure \ref{total}, a spectrum of
total energies in B1, B2, B3, S1X/Y and S2X/Y integrated
over a time window of 100 ns is shown.
The peaks at 15.6 and 18.5 MeV
are from $\pi^+ \rightarrow \mbox{e}^+ \nu$ and $\pi^+ \rightarrow \mu^+ \rightarrow \mbox{e}^+$ decays, respectively.
(The energy separation between the two peaks is smaller than
4.1 MeV because of saturation effects in the plastic scintillator.)
The vertical dashed
lines indicate the cut positions.
The ratio of the low energy events 
($E_{e^+} < 54$ MeV, including the $\pi^+ \rightarrow \mbox{e}^+ \nu$ tail) 
and the $\pi^+ \rightarrow \mbox{e}^+ \nu$ peak ($E_{e^+} > 54$ MeV) was 0.2, consistent 
with that obtained in the previous TRIUMF experiment 
\cite{oldneutrino} (before the optimization process described below).
At this stage, 
the major low-energy background in the background-suppressed spectrum came from
decay-in-flight
(DIF) of pions near the B3 counter, in which the muon from the $\pi^+ \rightarrow \mu^+ \nu$ decay
stopped in B3 and deposited the same kinetic energy as the
initial pion \cite{oldtriumf}.

\begin{figure}[htb]
\centering
\includegraphics*[width=8cm]{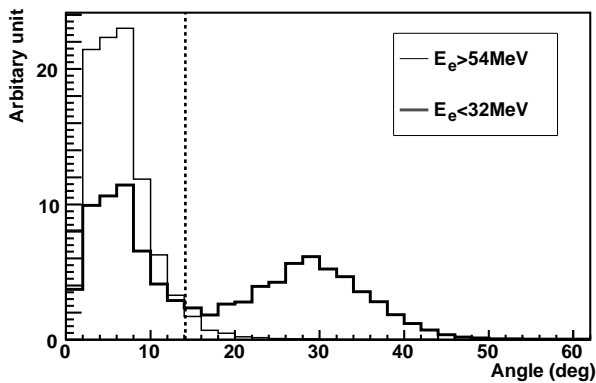}
\caption{Kink angles for events $E > 54$ MeV (thin) and
$E < 32$ MeV (thick).
The vertical dotted line at 14$^{\circ}$ indicates the cut position.}
\label{kink}
\end{figure}

The tracking detectors, S1X/Y and S2X/Y, allowed detection of a kink
 in the pion track when DIF happened upstream of the B3 counter.
For the remaining events, the ``pion'' direction near B3 
with respect to the beam direction
is plotted in Fig.\ref{kink}
for the regions $E_{e^+} > 54$ MeV (mostly $\pi^+ \rightarrow \mbox{e}^+ \nu$) and
$E_{e^+} <32$ MeV (mostly pion DIF events 
with a negligible $\pi^+ \rightarrow \mbox{e}^+ \nu$ tail contribution)
by thin and thick lines, respectively. 
The background was suppressed by another factor of two by
requiring the kink angle to be $< 14^{\circ}$.

Also, consistency tests of the ``pion'' and  positron tracks 
based on the closest approach of the two tracks in B3
provided an extra
handle for suppressing the pion DIF events.
An additional  suppression factor of three was obtained with a loss
of statistics of 30 \%.

\begin{figure}[htb]
\centering
\includegraphics*[width=8cm]{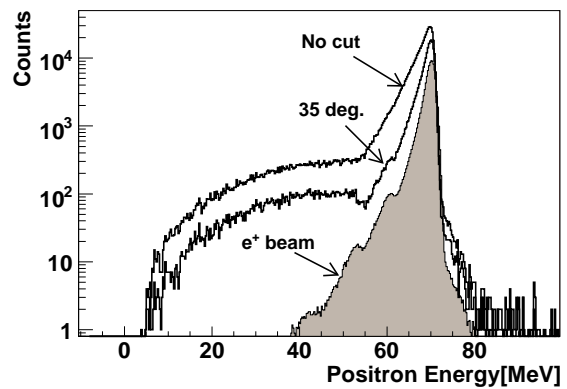}
\caption{
Positron spectra with (35$^{\circ}$) and without (No cut)
angle cuts together with the positron beam spectrum shifted by 2.5 MeV
(shaded).}
\label{angle}
\end{figure}

The cuts were optimized by minimizing the value 
$S = \sqrt{N_{<54 {\rm MeV}}} / N_{>54 {\rm MeV}}$,
where $N_{<54 {\rm MeV}}$ and $N_{>54 {\rm MeV}}$ are the numbers of
events below and above 54 MeV
in the positron energy spectrum, respectively.
The final background (including the low-energy tail of the
$\pi^+ \rightarrow \mbox{e}^+ \nu$ peak) to peak ratio was $N_{<54 {\rm MeV}} / N_{>54 {\rm MeV}}=0.068$
with $N_{>54 {\rm MeV}} = 4.8 \times 10^5$.
The positron energy spectrum 
 is shown as ``No cut'' in Fig. \ref{angle}.
\\

\subsection{Spectra with angle cut}

There was a strong correlation between the emission angles of
the decay positrons and
the amount of the
low energy tail due to shower leakage. 
As shown in Fig.\ref{angle}, the peak resolution was also
improved by 10 \% with the positron angle cut at 35$^{\circ}$
 with respect to the beam axis
($\sigma = 0.8$ \% (RMS) at 70 MeV with the 35$^{\circ}$ cut). 
This cut also contributed to a better peak-to-background ratio.
However, tighter cuts on the positron
emission angle worsened the $S$ value.
Using spectra with the angle cut
was effective only above 47 MeV
where the impact of the resolution was higher.

There was
a shoulder in the positron spectrum
approximately at 60 MeV that was enhanced with the angle cut.
Similar structures observed in the spectrum
of the 75-MeV/c beam positron
shown by the shaded histogram in Fig.\ref{angle}
were discussed in Ref.\cite{bump}.
(To compensate for the difference in the initial energies and the
additional energy losses of the beam positrons in B1, B2 and B3,
the beam-positron spectrum was shifted by 2.5 MeV 
to line up with the $\pi^+ \rightarrow \mbox{e}^+ \nu$ peak.)
The bumps (or shoulders) at 60 MeV and 53 MeV
in the beam positron spectrum
correspond to the primary peak energy minus one and 
two neutron separation
energies in $^{127}$I, respectively;
the loss of the energy observed in the NaI(T$\ell$) crystal is due to
low-energy neutrons produced in photo-nuclear reactions
 escaping from the crystal.
Since this 60 MeV shoulder in the suppressed spectrum is consistent with the
response function of the NaI(T$\ell$) crystal \cite{bump},
 it was treated as background
in the present study.
\\

\begin{figure}[htb]
\centering
\includegraphics*[width=8cm]{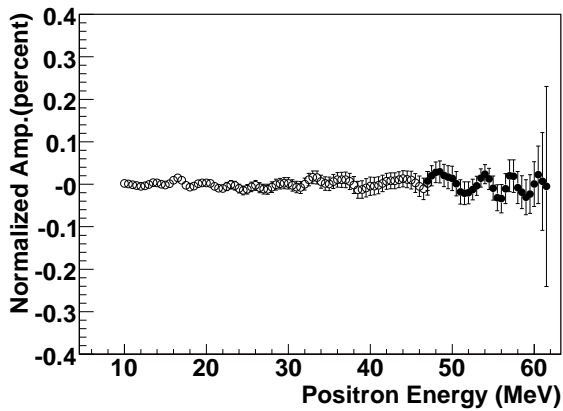}
\caption{Normalized amplitudes (\%) of potential peaks for the no-cut spectrum
(open circles) and for the 35$^{\circ}$ spectrum 
(closed circles).
}
\label{amplitude}
\end{figure}

\subsection{Fitting}

The search for extra peaks 
was conducted 
with 0.5 MeV steps
in the positron energy regions, $E_{e^+} = 10-47$ MeV for the spectrum
without the angle cut and  $E_{e^+} = 47-60$ MeV with the 35$^{\circ}$ cut
to take advantage of improved energy resolution 
obtained using the cut.
 The entire spectrum (9-50 MeV for the no-cut data and 9-62 MeV for the
data with the 35$^{\circ}$ cut) was fitted to a background function (described
below) plus a possible peak.

In order to minimize the effects of the
bumps in the response function,
the beam positron spectrum was subtracted from the
spectra before the fitting search.
 The normalization was done in such a way that the
60 MeV peak amplitude in the fit of the spectrum with the 35$^{\circ}$ cut was zero. 

The amplitude of the $\pi^+ \rightarrow \mu^+ \rightarrow \mbox{e}^+$ background, for which
the spectrum was obtained with a late time window (150--500 ns),
 was a free parameter in the fit.

The positrons from DIF muons emitted promptly following $\pi^+ \rightarrow \mu^+ \nu$ decays
(about one third of the total $\pi^+ \rightarrow \mu^+ \rightarrow \mbox{e}^+$ background) 
decreased with the pion decay time, 
and the $\pi^+ \rightarrow \mu^+ \rightarrow \mbox{e}^+$ spectrum obtained from the late time window
did not include this component.
The muon-DIF spectrum was obtained by applying a Lorentz transformation
for the muon kinetic energies 3.3--4.1 MeV to the $\pi^+ \rightarrow \mu^+ \rightarrow \mbox{e}^+$ spectrum
with acceptance corrections (described in the next section)
and muon polarization effects.
The resulting muon-DIF spectrum had a broad bump around 30--40 MeV,
extending near the $\pi^+ \rightarrow \mbox{e}^+ \nu$ peak. The amplitude of this component was a free
parameter of the fit.
 
In order to accommodate a slowly changing spectrum
mismatch and unspecified background, the amplitude and the decay constant of
an exponential function and an additional constant term were free
parameters of the fit.

 The template peak spectrum for each peak energy was obtained by applying the
same cuts used in the data analysis to the positrons which
were generated by a Monte Carlo (MC) simulation in the B3 counter 
with the observed pion stopping distribution.
Measurements using positron beams at various entrance
angles and energies into the NaI(T$\ell$) crystal
confirmed the validity of the MC line shapes including the effects of CsI
vetoing. Agreements between the data and MC in the peak shape
were within 10 \%.

Figure \ref{amplitude} shows normalized amplitudes of fitted peaks
for the spectra without the angle cut  (open circles)  and
with the 35$^{\circ}$ angle cut (closed circles).
The $\chi^2 / \mbox{DOF}$'s were 0.97 without the angle cut and 1.00 with the
angle cut for the fits without the extra peak.
\\
 
\subsection{Acceptance}

Since the low-energy peak amplitudes obtained were normalized 
to that of the 70 MeV peak, 
most acceptance effects
canceled to first order, especially those related to
the pion definition cuts.
There were, however, some energy-dependent effects in the cuts 
to be corrected for.
The acceptances were estimated based on MC calculations. The consistency
was tested to be within 3 \% by comparing the MC and experimental
$\pi^+ \rightarrow \mu^+ \rightarrow \mbox{e}^+$ spectra in the 10--50 MeV region with and without the background 
suppression cuts.

The fiducial cut increased the relative acceptance for 10 MeV
positrons by 5 \% (for no angle cut) with respect to 70 MeV positrons
due to multiple scattering effects. 
Energy leakage into the CsI crystals
for higher positron energy resulting in rejection of events
also increased the relative
acceptance of 10 MeV positrons by 15 \%. 

Because of larger scattering cross sections at lower energy,
even within a small path length in B3 (3 mm in depth),
lower energy positrons tended to have a larger total
energy deposit in the target, thus lower energy events
 looked more like $\pi^+ \rightarrow \mu^+ \rightarrow \mbox{e}^+$
decays, causing a 10 \% loss (no angle-cut data) 
in efficiency due to the total energy cut.
The largest energy-dependent effect was in the vertex consistency
requirement for
pion and positron tracks, which reduced the acceptance of low energy
positrons by 60 \%.
The combined acceptances for 10 MeV positrons with respect to 70 MeV
positrons were 45 \%  (35$^{\circ}$ data) and 42 \% (no cut).
\\

\section{Results}

\begin{figure}[htb]
\centering
\includegraphics*[width=8cm]{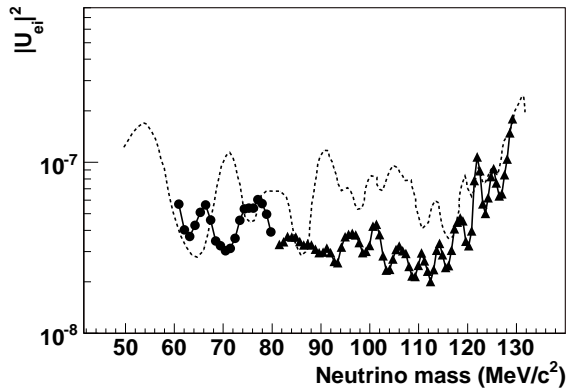}
\caption{Combined 90 \% C.L. upper limits obtained from the
 35$^{\circ}$ spectrum (circles)
and no-cut spectrum (triangles) 
together with the previous limits (dashed line) \cite{oldneutrino}.}
\label{uei}
\end{figure}

No significant peaks above statistical fluctuations were observed.
After correcting for the acceptance
and the helicity-suppression and phase-space terms,
the amplitudes and associated errors were converted to 90 \% C.L. 
upper limits on $|U_{ei}|^2$,
assuming a Gaussian probability distribution with a constraint
that the physical region of a peak area be positive.
Figure \ref{uei}  shows the combined results for the fits with
the 35$^{\circ}$ angle cut (below 80 MeV/c$^2$ in neutrino mass),
 and without the angle cut (above 80 MeV/c$^2$).
The region below 60 MeV/c$^2$ ($E_{e^+} >57$ MeV) was excluded 
in the plot because of
the strong bias caused by the background subtraction procedure.
For comparison, the 90 \% C.L. upper limits obtained in Ref. \cite{oldneutrino}
are also plotted by a dashed curve.
\\

\section{Conclusions}

The present experiment improved the upper limits
on the neutrino mixing matrix element $|U_{ei}|^2$
by a factor of
up to four in the mass region 68--129 MeV/c$^2$.
\\ 

\begin{acknowledgments}
This work was supported
 by the Natural Science and Engineering Research Council
and TRIUMF through a contribution from the National Research Council of Canada,
and by Research Fund for the Doctoral Program of Higher Education of China, and
partially supported by KAKENHI (21340059) in Japan,
One of the authors (M.B.) has been supported by US National Science
Foundation Grant Phy-0553611.
We are grateful to Brookhaven National Laboratory for
the loan of the crystals, and to the TRIUMF detector, electronics and DAQ groups for
their engineering and technical support.
\end{acknowledgments}


\begin{thebibliography}{9}
\bibitem{sterile} A. Kusenko, Phys. Rep. 481, 1 (2009) and references
therein.
\bibitem{numsm}A. Boyarsky, O. Ruchayskiy and M. Shaposhnikov,
Ann. Rev. Nucl. Part. Sci. 59 (2009) 191-214.
\bibitem{niigata}T. Asaka, S. Eijima and H. Ishida,
J. High Energy Phys. 1104, 011 (2011).
\bibitem{oldtriumf} D.I. Britton $et~al.$, Phys. Rev. Lett. 68,
 3000 (1992);
and D.I. Britton $et~al.$, Phys. Rev. D49, 28 (1994).
\bibitem{oldpsi}G. Czapek $et~al.$, Phys. Rev. Lett. 70,
17 (1993).
\bibitem{pdg2010} K. Nakamura $et~al.$ (Particle Data Group),
J. Phys. G37, 075021 (2010).
\bibitem{oldneutrino} D.I. Britton $et~al.$,
 Phys. Rev. D46, R885 (1992).
\bibitem{pienu} PIENU experiment, TRIUMF Proposal S1072, (2005).
\bibitem{m13}  A. Aguilar-Arevalo $et~al.$,
Nucl. Instrum. Method A609, 102 (2009).
\bibitem{bnl} G. Blanpied $et~al.$, Phys. Rev. Lett. {\bf 76} (1996) 1023.
\bibitem{e949} I-H. Chiang $et~al.$, IEEE {\bf NS-42} (1995) 394.
\bibitem{bump} A. Aguilar-Arevalo $et~al.$,
Nucl. Instrum. Method A621, 188 (2010).
\end{thebibliography}
\end{document}